\documentclass[aps,prl,twocolumn,nofootinbib]{revtex4-2}

\usepackage{amsfonts}
\usepackage{amsmath}
\usepackage[normalem]{ulem}
\usepackage{amssymb}
\usepackage{mathtools}
\usepackage{graphicx}
\usepackage{enumerate}
\usepackage{wrapfig}
\usepackage{color}
\usepackage{yhmath}
\usepackage{mathrsfs}
\usepackage{bbm}
\usepackage{commath}
\usepackage[hidelinks]{hyperref}
\hypersetup{
     colorlinks   = true,
     citecolor    = blue,
     linkcolor    = blue
}
\usepackage{subfigure}
\usepackage{mathptmx}
\usepackage{times}
\usepackage{ulem}

\DeclareMathAlphabet{\pazocal}{OMS}{zplm}{m}{n}

\DeclareMathOperator{\tr}{tr}

\DeclareMathSymbol{\Alpha}{\mathalpha}{operators}{"41}
\DeclareMathSymbol{\Beta}{\mathalpha}{operators}{"42}
\DeclareMathSymbol{\Epsilon}{\mathalpha}{operators}{"45}
\DeclareMathSymbol{\Zeta}{\mathalpha}{operators}{"5A}
\DeclareMathSymbol{\Eta}{\mathalpha}{operators}{"48}
\DeclareMathSymbol{\Iota}{\mathalpha}{operators}{"49}
\DeclareMathSymbol{\Kappa}{\mathalpha}{operators}{"4B}
\DeclareMathSymbol{\Mu}{\mathalpha}{operators}{"4D}
\DeclareMathSymbol{\Nu}{\mathalpha}{operators}{"4E}
\DeclareMathSymbol{\Omicron}{\mathalpha}{operators}{"4F}
\DeclareMathSymbol{\Rho}{\mathalpha}{operators}{"50}
\DeclareMathSymbol{\Tau}{\mathalpha}{operators}{"54}
\DeclareMathSymbol{\Chi}{\mathalpha}{operators}{"58}
\DeclareMathSymbol{\omicron}{\mathord}{letters}{"6F}

\def\XXint#1#2#3{{\setbox0=\hbox{$#1{#2#3}{\int}$}
\vcenter{\hbox{$#2#3$}}\kern-.5\wd0}}

\makeatletter
\DeclareFontFamily{OMX}{MnSymbolE}{}
\DeclareSymbolFont{MnLargeSymbols}{OMX}{MnSymbolE}{m}{n}
\SetSymbolFont{MnLargeSymbols}{bold}{OMX}{MnSymbolE}{b}{n}
\DeclareFontShape{OMX}{MnSymbolE}{m}{n}{
    <-6>  MnSymbolE5
   <6-7>  MnSymbolE6
   <7-8>  MnSymbolE7
   <8-9>  MnSymbolE8
   <9-10> MnSymbolE9
  <10-12> MnSymbolE10
  <12->   MnSymbolE12
}{}
\DeclareFontShape{OMX}{MnSymbolE}{b}{n}{
    <-6>  MnSymbolE-Bold5
   <6-7>  MnSymbolE-Bold6
   <7-8>  MnSymbolE-Bold7
   <8-9>  MnSymbolE-Bold8
   <9-10> MnSymbolE-Bold9
  <10-12> MnSymbolE-Bold10
  <12->   MnSymbolE-Bold12
}{}

\newcommand{\ignore}[1]{}
\newcommand{\nobibentry}[1]{{\let\nocite\ignore\bibentry{#1}}}

\newcommand{\bibfnamefont}[1]{#1}
\newcommand{\bibnamefont}[1]{#1}
\newcommand{\ket}[1]{\left\vert#1\right\rangle}
\newcommand{\bra}[1]{\left\langle#1\right\vert}

\newcommand{\bea}{\begin{eqnarray}}
\newcommand{\eea}{\end{eqnarray}}

\renewcommand*{\thefootnote}{\fnsymbol{footnote}}

\begin{document}

\title{Thermodynamics of the Page curve in Markovian open quantum systems}

\author{Jonas Glatthard}
\affiliation{School of Physics and Astronomy, University of Nottingham, Nottingham NG7 2RD, UK}
\email{jonas.glatthard@nottingham.ac.uk}

\begin{abstract}
Typically, the von Neumann entropy of a subsystem increases until it plateaus at the thermal value. Under some circumstances, however, the intermediate value can dwarf the final value, even if the subsystem starts in a pure state. A famous example in the context of the black hole information paradox is the entropy of the Hawking radiation, where this behaviour is dubbed the Page curve. More generally, this is the case for excited systems weakly coupled to cold reservoirs. Here we study the entropy dynamics for Lindbladian evolution, i.e. open quantum systems in weak contact with Markovian reservoirs. This allows us to study the non-equilibrium thermodynamics of the subsystem entropy decrease and link it to Landauer’s principle: the entropy decrease must be accompanied by a heat flow out of the system. We give an analytic expression of the entanglement dynamics for a decaying excitation in a two-level system and study it under equilibration of a localised oscillator. In both cases the Page time occurs when half the initial energy has left the system.
\end{abstract}

\maketitle

\noindent \textit{Introduction.} Entanglement is at the forefront of research in a wide range of subfields of physics, from quantum information to condensed matter and high-energy physics. An important tool to quantify the entanglement of subsystems of an overall system in a pure state is the entanglement entropy, i.e. the von Neumann entropy of the subsystem, which equals that of its complement. It shows rich characteristics in quantum many-body systems such as a ground state area law for gapped systems, logarithmic behavior for critical systems and a volume law for exited states \cite{calabrese2009, eisert2010, lydzba2020, cirac2021, bianchi2021, bianchi2022}. Furthermore, the entanglement structure of interacting quantum systems has become an important foundation on which numerical approximations are based. For example, matrix product states have become state-of-the-art for the numerical simulation of quantum many-body systems \cite{schollwock2011} and open quantum systems \cite{strathearn2018}.
Under unitary dynamics, the typical behavior of the entanglement entropy of a subsystem is to increase until it plateaus at an equilibrium value \cite{calabrese2005,ho2017,kim2013},
often given by the eigenstate thermalisation hypothesis \cite{deutsch1991,srednicki1994,alessio2016,gogolin2016}.

An important deviation from the typical entanglement dynamics is the entropy of Hawking radiation \cite{almheiri2021}. This radiation is emitted by black holes due to quantum fields, has a thermal spectrum and leads to evaporation of the black hole \cite{hawking1975}. It associates a temperature and entropy \cite{bekenstein1972} to black holes and makes them thermodynamic objects \cite{wald2001}. At the end of Hawking's process stands flat space filled with the emitted thermal radiation, seemingly erasing all information contained in the black hole. This is the so-called black hole information paradox. If the overall process is unitary, as would be expected from a fully quantum treatment, information must be conserved and the final entanglement entropy of the radiation zero \cite{page1993}. The entropy is then expected to first increase, as in Hawking's semiclassical calculation, but, when roughly half the black hole mass has radiated away, start to decrease back to zero--this is the so-called Page curve. Recently, there has been substantial progress in recovering the Page curve for evaporating black holes using holography \cite{ryu2006prl,ryu2006jhep} and the semiclassical gravitational path integral \cite{almheiri2019, penington2020}.

One can view the radiation fields as acting as a bath on the black hole, which is then an open system \cite{bp, weiss1999}.  The entanglement of an open quantum system with its environment is under active study \cite{hilt2009, benatti2010, morozov2012, salamon2017, bernardo2021, roszak2021, ptaszynski2022, zhan2021, salamon2023, ptaszynski2024, kobayashi2024}.
A number of works have recently highlighted, that a Page-curve-like entanglement dynamics is common for quantum systems coupled to a reservoir \cite{kehrein2024, glatthard2024a, saha2024}. Ref.~\cite{kehrein2024} points out that it can arise if the dissipative dynamics drives the system into a low-dimensional subspace.
Ref.~\cite{glatthard2024a} argues that this is generically realised by weak coupling to a cold reservoir. More specifically, after preparing the open system in a non-equilibrium pure state, the interaction entangles system and environment, but in the long time the open system approaches the mean-force Gibbs state \cite{trushechkin2022}, carrying low entropy. At zero temperature the overall system is pure, and the environment, playing the role of the Hawking radiation, follows the same entanglement dynamics.
In turn, open system techniques have been used to study black hole physics \cite{su2021, li2024, geng2024}.

The studies \cite{kehrein2024, glatthard2024a, saha2024} worked with exact methods, either with solvable systems \cite{caldeira1983path} or numerical simulation
\cite{tanimura2020}. The current work is motivated by extending the analysis using standard tools for the study of open quantum systems, in particular Markovian master equations. Those are effective equations for the state of the open system.  Such master equations are one of the cornerstones of quantum thermodynamics \cite{alicki1979, palao2001,kosloff2014, vinjanampathy2016, binder2018}, i.e. the study of thermodynamic processes in the quantum regime. Beyond simplifying the time evolution, this has thus the advantage that we can link the entropy dynamics to thermodynamics. In particular, the so-called \emph{global master equation} \cite{gonzalez2017, hofer2017, cattaneo2019} not only leads to Markovian dynamics \cite{davies1974,lindblad1976,gks1976}, but to a well-defined non-equilibrium thermodynamics \cite{spohn1977, spohn1978, dann2021, glatthard2024c}. Crucially, it describes equilibration of the open systems at the temperature of the bath \cite{spohn1977} and fulfills an entropy balance equation \cite{spohn1978}. This means, as we will see, equilibration at absolute zero enforces the entropy decrease to zero at long times. Furthermore, the entropy balance equation for the global master equation implies that the entropy decrease must be accompanied by a heat flow out of the system. We will argue that this is an instance of Landauer's principle \cite{landauer1961}.

In the following, we outline the essentials of the global master equation, its entropy dynamics and non-equilibrium thermodynamics. We then turn to two specific examples of Page-curve-like entanglement dynamics for Markovian open systems. The first is particularly simple and illustrative: a decaying excitation in a two-level system. The simplicity of this example allows us to write down an analytical expression for the entanglement dynamics. \\

\noindent \textit{Entropy under the global master equation.} Microscopically, an open quantum system (`S') is described together with its environment, or bath (`B'), via a Hamiltonian of the form $\pmb H = \pmb H_S + \pmb H_B + \pmb H_I$. Master equations are effective equations involving only degrees of freedom of the system, by tracing out the environment. Most commonly, one considers the case of a factorised initial state $\pmb \rho \otimes \pmb \tau_B$, where $\pmb \rho$ is an arbitrary state of the system and $\pmb \tau_B$ a thermal state of the bath. Invoking the Born (weak coupling), Markov and secular approximation one arrives at a Markovian master equation, $\pmb{\dot \rho} = \mathcal L (\pmb \rho)$, in GKLS-form \cite{lindblad1976,gks1976} (after Gorini, Kossakowski, Lindblad and Sudarshan). A superoperator $\mathcal L$ in GKLS-form contains, next to a Hamiltonian part, a number of positively weighed terms of form
\begin{equation}
 \mathcal{D}[\pmb A](\cdot)= \pmb A \cdot \pmb A^\dagger - \frac{1}{2} \{\pmb A^\dagger \pmb A, \cdot \},
\end{equation}
where the jump operator $\pmb A$ is an arbitrary, not necessarily Hermitian operator on the system. This form guarantees complete positivity, is trace-preserving, and the resulting evolution is a semigroup.

More specifically, following the procedure outlined above, one gets a number of thermodynamically appealing properties. For conciseness, we give here the expressions for an interaction (`I') term of form $\pmb H_I = \pmb S \otimes \pmb B$, although the later results, by linearity, hold generally for sums of such terms. Following the procedure, one arrives at a master equation in Davies form \cite{davies1974} \footnote{The Hamiltonian part could additionally include a so-called Lamb shift. It can however be canceled via an renormalisation counter-term in the total Hamiltonian \cite{correa2023}.} 
\begin{equation} \label{eq:Davies}
 \pmb{\dot \rho}= - \rm{i} \left[\pmb H_S, \pmb \rho \right] + \sum_\omega \gamma_\omega \mathcal{D}[\pmb A_\omega](\pmb \rho),
\end{equation}
where the sum runs over the Bohr frequencies $\omega$ of $\pmb H_S$ and the jump operators $\pmb{A}_{\omega}$ are constructed such that $ \pmb{S} = \sum\nolimits_\omega \pmb{A}_{\omega} $ and $[\pmb{H}_S,\pmb{A}_\omega] = - \omega\,\pmb{A}_\omega$. The rates $\gamma_\omega$ are given through the two-point function of $\pmb B$ on the bath and fulfill detailed balance $\gamma_{-\omega}/\gamma_\omega = \rm{e}^{-\beta \omega}$, where $\beta$ is the inverse temperature $T$ of the bath. Here, and in what follows, we work in units such that $\hbar = k_B = 1$. This master equation is often called the global master equation. Alternatively, instead of a derivation from a microscopic system-bath model, such a master equation uniquely follows from a number of thermodynamics-inspired postulates \cite{dann2021}.

Detailed balance guarantees that the Gibbs state at temperature $T$ is a steady state of Eq.~\eqref{eq:Davies}. Moreover, if only operators proportional to the identity commute with all $\pmb H_S, \pmb A_\omega$, it is the only steady state, approached by all $\pmb \rho$ \cite{spohn1977}, i.e.
\begin{equation} \label{eq:ss}
\pmb \rho \underset{t \rightarrow \infty}{\rightarrow} \pmb \tau \coloneq \rm{e}^{-\beta \pmb H_S}/\tr\left(\rm{e}^{-\beta \pmb H_S}\right).
\end{equation}
The global master equation therefore describes equilibration of a weakly coupled Markovian open quantum system.

Moreover, the von Neumann entropy,
\begin{equation} \label{eq:SvN}
	S = - \tr \left( \pmb \rho \log \pmb \rho \right),
\end{equation}
follows a balance equation under  Eq.~\eqref{eq:Davies}, taking a form well known from non-equilibrium thermodynamics \cite{spohn1978}, i.e.
\begin{equation} \label{eq:be}
	\dot S = \beta \mathcal{\dot Q} + \sigma.
\end{equation}
Here $\mathcal{\dot Q}$ is the heat entering the system
\begin{equation}
	\mathcal{\dot Q} = \tr \left( \pmb H_S \mathcal L (\pmb \rho) \right),
\end{equation}
and $\sigma$ the entropy production
\begin{equation}
	\sigma = - \dot S(\pmb \rho | \pmb \tau),
\end{equation}
with the quantum relative entropy $S(\pmb \rho | \pmb \rho') \coloneqq \tr\,\pmb{\rho}(\log{\pmb{\rho}}-\log{\pmb{\rho'}})$. The latter is non-negative and decreasing under the evolution~\eqref{eq:Davies}. We therefore have $\sigma \ge 0$. Although non-symmetric, the relative entropy is often used as a notion of distance between states as it vanishes iff $\pmb \rho = \pmb \rho'$. When the system is coupled to multiple baths, the steady-state version of Eq.~\eqref{eq:be} becomes the Clausius inequality \cite{glatthard2024b}.

We can now relate the entropy dynamics to thermodynamics. After initialising the system in a pure state, the interaction with a bath at zero temperature can generate a high amount of entanglement entropy, as we will see in the examples below. However, the final value of the entropy is given by the entropy of the steady state, Eq.~\eqref{eq:ss}. For $T=0$ this is the local ground state, which, for non-degenerate Hamiltonians, has $S=0$. In-between, we thus require $\dot S < 0$. As $\sigma \ge 0$, Eq.~\eqref{eq:be} states that $\dot S < 0$ requires $\mathcal{\dot Q} < 0$, i.e.
\begin{equation}
    \dot S < 0 \Rightarrow \mathcal{\dot Q} < 0.
\end{equation}
The entropy decrease therefore must be accompanied by a heat flow out of the system. In other words, it occurs only for exothermic processes.

We can connect this condition to a fundamental principle of information thermodynamics. For that, we identify the heat flow into the bath as $\mathcal{\dot Q}_B = -\mathcal{\dot Q}$. This assumes that no energy is stored at the system-bath interface, which is valid under the weak coupling assumption \cite{dann2021}. We can then restate Eq.~\eqref{eq:be} as
\begin{equation} \label{eq:mlb}
    -T \dot S \le \mathcal{\dot Q}_B, 
\end{equation}
this is the differential version of Landauer's principle for Markovian open systems. Landauer's principle states that erasing information has a fundamental heat cost \cite{landauer1961}, i.e. the change of entropy of a system $\Delta S$ is bounded by the heat $\Delta \mathcal{Q}_B$ dissipated to the environment as \cite{timpanaro2020}
\begin{equation}
-T \Delta S \le \Delta \mathcal{Q}_B.
\end{equation}
For $\Delta S<0$ Landauer's principle bounds the minimum heat cost necessary to purify a system. We note that under the Markovian approximation, Eq.~\eqref{eq:mlb} is valid also for $t>0$, when the system is entangled with the bath, and the bath departs from the thermal state, unlike in the typical setup for Landauer's principle \cite{timpanaro2020}. It is the structure of the master equation~\eqref{eq:Davies}, which allows the formulation of a differential version of Landauer's principle.\\

\begin{figure}[t]
\centering
\includegraphics[width=0.45\textwidth]{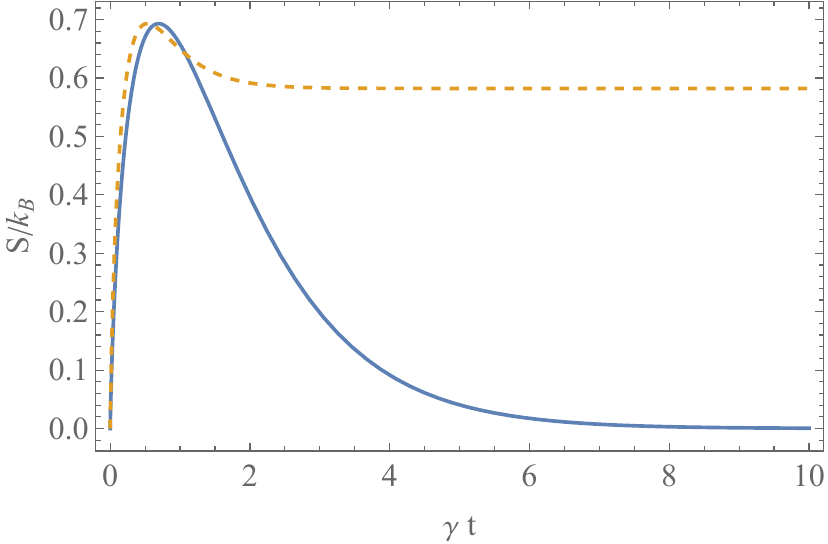}
\caption{The entropy dynamics of an excited two-level system in a zero temperature environment (solid blue) as described by the master equation~\eqref{eq:eom-tls} follows the Page curve. As the excited state is pure, the entropy starts at zero. The interaction with the environment generates entanglement entropy as the excitation decays. When half the initial energy left the system, it passes through the fully mixed state, which carries the maximal entropy. In the long time, the master equation brings the system to its ground state and thus the entropy back to zero. An analytical expression of the curve is given in Eq.~\eqref{eq:analytical}. The overall system-plus-environment state is and remains pure, thus the entropy of the environment follows the same evolution.
In contrast, at high temperatures ($T=1$, dashed orange) the entropy dynamics behaves more typical and saturates at a high value. In both cases $\epsilon_0=1, \gamma=0.01$.}
\label{fig1}
\end{figure}

\begin{figure}[t]
\centering
\includegraphics[width=0.45\textwidth]{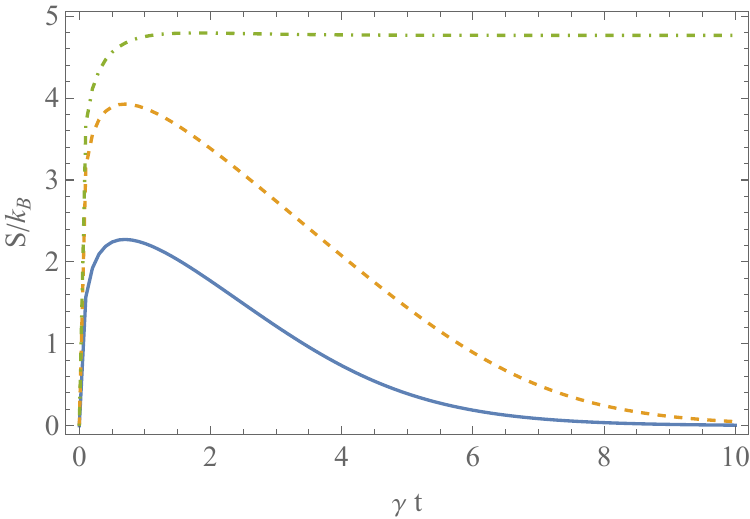}
\caption{The entropy dynamics of a localised oscillator ($\delta=10^{-3}$, solid blue) in a zero temperature environment as described by the master equation~\eqref{eq:eom-ho} follows the Page curve. The more localised ($\delta=10^{-4}$, dashed orange) the initial state, the higher the intermediate entropy. As in the previous example, the Page time is reached when half the energy left the system. Regardless of the initial state, the master equation brings the system to its ground state with zero entropy. Again, at hight temperatures ($T=10$, dot-dashed green) the entropy dynamics behaves typically. In all cases $\omega_0=1, \gamma=0.01$.}
\label{fig2}
\end{figure}

\noindent \textit{Examples.} We now turn to examples. The first example is a quantum two-level system, evolving under the master equation
\begin{equation} \label{eq:eom-tls}
 \pmb{\dot \rho}= - \rm{i} \left[\frac{\epsilon_0}{2} \pmb \sigma_z, \pmb \rho \right] + \gamma_- \mathcal{D}[\pmb \sigma_-](\pmb \rho) + \gamma_+ \mathcal{D}[\pmb \sigma_+](\pmb \rho),
\end{equation}
where $\epsilon_0$ is the energy gap, the $\pmb \sigma_i$ are the Pauli matrices and $\pmb \sigma_\pm = (\pmb \sigma_x \pm \rm{i} \pmb \sigma_y)/2$ the raising/lowering operators. Such jump operators arise from an interaction with $\pmb S \propto \pmb \sigma_x$, as outlined below Eq.~\eqref{eq:Davies}. The rates are given by $\gamma_\pm = \mp \gamma \epsilon_0 \left(1 + (\rm{e}^{\mp \beta \epsilon_0}-1)^{-1} \right)$, where $\gamma$ is the system-bath coupling strength. At $T=0$ the last term in Eq.~\eqref{eq:eom-tls} vanishes.

We initialise the system in the initial state $\pmb \rho_0 = \ket{1} \bra{1}$ and let it evolve according to Eq.~\eqref{eq:eom-tls} at $T=0$, i.e. for an environment in its ground state $\pmb \rho_0 = \ket{\Omega} \bra{\Omega}$. The energy of the system (relative to the ground state energy $E_0$) will then exponentially decay with the excited state population
\begin{equation}
 E(t) - E_0 \propto \bra{1} \pmb \rho(t) \ket{1} = \rm{e}^{-\gamma t}, 
\end{equation}
while the off-diagonals in the energy-basis remain zero.
Together with the ground state population, via unit trace, we can calculate the von Neumann entropy with Eq.~\eqref{eq:SvN} to obtain the analytical expression
\begin{equation} \label{eq:analytical}
 S(t) = \gamma t \rm{e}^{-\gamma t} - (1-\rm{e}^{-\gamma t}) \log(1-\rm{e}^{-\gamma t}).
\end{equation}
As the overall initial state $\ket{1} \otimes \ket{\Omega}$ is pure and the overall dynamics Hermitian this is the entanglement entropy. It starts at zero, reaches a maximum of $S_*=ln(2)$ at $t_*=ln(2)/\gamma$, and then decays back to zero, in accordance with Eq.~\eqref{eq:ss}. It thus follows the Page curve. The time it reaches its maximum and turns over, i.e. the Page time $t_*$, is also the time when half of the initial energy has decayed. During the whole evolution heat flow out of the system, in accordance with our discussion in previous sections. This entanglement dynamics is illustrated in Fig.~\ref{fig1}. We compare it to the entropy dynamics at $T>0$, which behaves more typically and saturates at the thermal value much larger than zero.

We now turn to our second example, the dissipative quantum oscillator. It evolves under the master equation
\begin{equation} \label{eq:eom-ho}
 \pmb{\dot \rho}= - \rm{i} \left[\omega_0 \pmb a^\dagger \pmb a, \pmb \rho \right] + \gamma_- \mathcal{D}[\pmb a](\pmb \rho) + \gamma_+ \mathcal{D}[\pmb a^\dagger](\pmb \rho),
\end{equation}
where $\omega_0$ is the frequency and $\pmb a, \pmb a^\dagger$ are the lowering/raising operators. Such jump operators arise from an interaction $\pmb S \propto \pmb a^\dagger + \pmb a$. The rates are given by $\gamma_\pm = \mp \gamma \omega_0 \left(1 + (\rm{e}^{\mp \beta \omega_0}-1)^{-1} \right)$ and, as in the previous example, the last term in Eq.~\eqref{eq:eom-ho} vanishes at $T=0$. As Eq.~\eqref{eq:eom-ho} is \textit{Gaussianity}-preserving \cite{ferraro2005}, we work with Gaussian states. To do so, we express the state in terms of its quadrature variances $ \sigma_{xx} = \langle \pmb{x}^2 \rangle $, $ \sigma_{xp} = \frac12\langle \pmb{x}\pmb{p} + \pmb{p}\pmb{x} \rangle$ and $ \sigma_{pp} = \langle \pmb{p}^2 \rangle$, for $\pmb{x},\pmb{p} = (\pmb a^\dagger \pm \pmb a)/\sqrt{2}$ ($m=\omega_0=1$). We collect the covariances in the matrix
\begin{equation}
    \mathsf{\Sigma} = \begin{pmatrix} \sigma_{xx} & \sigma_{xp} \\ \sigma_{xp} & \sigma_{pp} \end{pmatrix}.
\end{equation}
For vanishing first moments this fully characterises Gaussian states of the oscillator. The von Neuman entropy of Gaussian states can be calculated as \cite{demarie2012}
\begin{equation} \label{eq:SG}
    S = \left(\lambda + \frac{1}{2}\right) \log_2 \left( \lambda + \frac{1}{2}\right)-\left(\lambda - \frac{1}{2}\right) \log_2 \left( \lambda - \frac{1}{2}\right),
\end{equation}
where $\lambda$ is the symplectic eigenvalue, i.e. the absolute value of the eigenvalues $\{i \lambda , - i \lambda \}$ of $\mathsf{\Sigma} \mathsf{\Omega}$, with the symplectic matrix 
\begin{equation}
\mathsf{\Omega} = \begin{pmatrix} 0 & 1 \\ -1 & 0 \end{pmatrix}.
\end{equation}

We illustrate the entropy dynamics of the dissipative oscillator in Fig.~\ref{fig2}. We initialise the system in a squeezed vacuum, i.e. $\sigma_{xx}=\delta$, $\sigma_{pp}=0.5^2/\delta$, $\sigma_{xp}=0$, an the bath at $T=0$. As the initial state is pure, the entanglement entropy starts at zero, reaches a maximum and then decays back to zero. The energy again decays exponentially as $E(t) - E_0 \propto \tr \left(\pmb a^\dagger \pmb a \pmb \rho(t) \right) \propto \rm{e}^{-\gamma t}$, the process is thus exothermic. As in the previous example, the time it reaches a maximum is the time at which half the initial energy has left the system, taking the value $t_*=ln(2)/\gamma$. The more localised the initial state, i.e. smaller $\delta$, the higher the maximum entropy $S_*$. In contrast, for large bath temperatures we get the typical entropy dynamics.\\

\noindent \textit{Conclusion.} In this letter we studied the Page-curve-like entanglement dynamics of open quantum systems using Markovian master equations. In particular, we saw that the behaviour of full system-bath models in Ref.~\cite{glatthard2024a} can be reproduced under this common approximation in the study of open quantum systems.
Moreover, the structure of the global master equation allowed us to connect the entropy dynamics to thermodynamics. In particular, the final entropy is given by the thermal state and the entropy balance equation \eqref{eq:be} implies that the decrease of entropy only takes place when heat flows out of the system. We can see this as an instance of Landauer's principle, which gives a minimum heat cost to purify a system.
Finally, we illustrated our discussion by two examples, a decaying two-level system and an equilibrating oscillator. In both cases the Page time corresponds to the time where half of the energy left the system. In particular the first example is simple and illustrative, the exponential decay of the excitation leads to a simple analytical formula for the entanglement entropy dynamics, 
 Eq.~\eqref{eq:analytical}.

In our analysis, the final value of the entropy is given by the steady state of the master equation and the initial value by the initial state. It might be of interest to find bounds on the intermediate time entanglement entropy. This could help characterise initial states and dissipation channels which generate a high amount of entanglement.
Also, the Page time occurring when half of the energy left the system reminisces the Page time of black holes, which ensues when roughly half of the initial mass has radiated away. One could study under what conditions this correspondence happens. One commonality in both our examples is that the systems lose their initial energy exponentially.
Further, it would be interesting to study whether a connection of the entropy dynamics to thermodynamics can be made outside the context of the global master equation.\\

\noindent \textit{Acknowledgments.} The author gratefully acknowledges
useful discussions with Andrew D. Armour, Mark T. Mitchison and Harry J. D. Miller. This work is supported by a Leverhulme Trust Research Project Grant (RPG-2023-177).\\

\renewcommand{\thefootnote}{\fnsymbol{footnote}}

\bibliographystyle{apsrev4-2}
\bibliography{references}

\end{document}